\documentclass{article}
\usepackage{spconf,amsmath,graphicx,hyperref}
\usepackage{booktabs}
\usepackage[dvipsnames]{xcolor}
\usepackage{multirow}
\usepackage{caption}
\usepackage{subcaption}

\title{Evaluating Latent Space Structure in Timbre VAEs: \\A Comparative Study of Unsupervised, Descriptor-Conditioned, and Perceptual Feature-Conditioned Models}
\name{Joseph M. Cameron, Alan F. Blackwell\thanks{The authors thank and acknowledge the Department of Computer Science, University of Cambridge \& the UKRI EPSRC for funding this research.}}
\address{Department of Computer Science \& Technology, University of Cambridge\\Cambridge, United Kingdom}
\begin{document}
%
\maketitle
\begin{abstract}
We present a comparative evaluation of latent space organization in three Variational Autoencoders (VAEs) for musical timbre generation: an unsupervised VAE, a descriptor-conditioned VAE, and a VAE conditioned on continuous perceptual features from the AudioCommons timbral models. Using a curated dataset of electric guitar sounds labeled with 19 semantic descriptors across four intensity levels, we assess each model’s latent structure with a suite of clustering and interpretability metrics. These include silhouette scores, timbre descriptor compactness, pitch-conditional separation, trajectory linearity, and cross-pitch consistency. Our findings show that conditioning on perceptual features yields a more compact, discriminative, and pitch-invariant latent space, outperforming both the unsupervised and discrete descriptor-conditioned models. This work highlights the limitations of one-hot semantic conditioning and provides methodological tools for evaluating timbre latent spaces, contributing to the development of more controllable and interpretable generative audio models.
\end{abstract}
%


\section{Introduction}
\label{sec:intro}

Controlling musical timbre with generative models is a key challenge in machine learning for audio. Variational Autoencoders (VAEs) are widely used in this domain due to their ability to learn latent spaces that support interpolation and continuous control \cite{Kingma_VAE_2014,Esling2018GenerativeTimbreSpaces}. However, the structure and interpretability of these latent spaces, particularly in relation to timbral descriptors, remain poorly understood. This limits their applicability in real-time, human-centered audio interfaces \cite{Caillon2020TimbreLatentSpace}.

Recent research has experimented with conditioning VAEs using both high-level semantic labels and low‑level acoustic features. For instance, Yutani et al. developed a wavetable synthesis model using a CVAE conditioned on semantic descriptors such as `bright', `warm', and `rich' for intuitive timbre control \cite{Yutani2024CVAEWavetable}. Roche et al. introduced a VAE that incorporates a perceptual regularization loss informed by listener ratings across timbral dimensions like `metallic' or `evolving', enabling controllable timbre transformations \cite{Roche2021PerceptualTimbreVAE}. Additionally, Natsiou et al. proposed a VAE trained on low-level acoustic features like spectral centroid and attack time to align latent representations with human perception of timbre \cite{Natsiou2023InterpretableTimbreVAE}.

Yet, no comprehensive study has systematically compared how such semantic versus perceptual conditioning strategies affect latent-space structure and interpretability for timbre generation and manipulation. Specifically, it remains unclear which VAE architectures best support consistent, discriminative, and controllable latent organization, qualities crucial for interactive audio applications.

This paper presents a comparative analysis of three VAE models trained on a curated dataset of electric guitar sounds with annotated timbre descriptors: (1) an unsupervised VAE, (2) a descriptor-conditioned VAE using one-hot semantic labels, and (3) a perceptual feature-conditioned VAE using continuous AudioCommons timbral features \cite{Pearce_AudioCommonsTimbreDescriptors_2017,Pearce_AudioCommonsTimbralModels_2019}. We evaluate each model’s latent space using both standard clustering metrics and novel interpretability measures tailored to timbre, including within-pitch separation, magnitude progression linearity, and cross‑pitch consistency.

Our results show that perceptual conditioning yields significantly more structured and pitch-invariant latent spaces, outperforming both the unsupervised and semantic descriptor conditioned variants. These findings provide evidence that perceptually grounded conditioning supports more usable latent representations for musical control tasks, and offer practical evaluation tools for future generative audio models.


\section{Related Work}
\label{sec:RelatedWork}

Generative models have become increasingly central in computational audio synthesis \cite{Esling2018GenerativeTimbreSpaces}, particularly for representing and manipulating musical timbre \cite{Caillon2020TimbreLatentSpace}. Among these, Variational Autoencoders (VAEs) \cite{Kingma_VAE_2014} have gained popularity due to their ability to learn continuous and interpretable latent spaces that support interpolation and continuous control. Recent work has explored conditioning VAEs to enhance control and improve disentanglement. Conditioning strategies include symbolic labels such as instrument class \cite{Bitton2019ModulatedVAE}, musical pitch \cite{Luo2019DisentangleTimbrePitch}, or expressive descriptors \cite{Yutani2024CVAEWavetable}. However, semantic labels can be noisy, subjective, and/or coarse-grained, which may hinder their effectiveness for fine-grained timbre control \cite{Luo2019DisentangleTimbrePitch}.

To address this, other approaches have used continuous perceptual features derived from audio analysis toolkits such as Timbre Toolbox \cite{peeters_descriptors_2011} or the AudioCommons project \cite{Pearce_AudioCommonsTimbreDescriptors_2017,Pearce_AudioCommonsTimbralModels_2019}. These features include timbral dimensions like brightness, roughness, or warmth, which have been shown to align well with human perceptual ratings \cite{Eerola2013MusicalCuesEmotion}. Roche et al. proposed a VAE with perceptual regularization based on human annotations \cite{Roche2021PerceptualTimbreVAE}, while Natsiou et al. trained a VAE to reconstruct and organize audio samples based on low-level acoustic descriptors \cite{Natsiou2023InterpretableTimbreVAE}.

Despite these developments, few studies have directly compared how different conditioning strategies impact latent space organization, interpretability, and cross-pitch consistency noted in pitch-timbre disentanglement \cite{Luo2019DisentangleTimbrePitch}. Furthermore, existing evaluations often rely on standard clustering metrics \cite{Manning_PurityScore_2008,Rousseeuw_SilhouetteScore_1987} without addressing musical domain-specific needs like trajectory linearity or timbre magnitude progression. This paper fills that gap by combining standard metrics with tailored interpretability measures to systematically compare three VAE variants for musical timbre modeling.


\section{Methodology}
\label{sec:Methodology}

To enable a valid and unbiased comparison of latent space structures, all three models in this study were trained under identical experimental conditions. Each model was trained on the same curated dataset, with consistent input representations, architecture capacity, training schedules, and optimization settings. This experimental design isolates the effect of the conditioning strategy on the resulting latent space structure, enabling direct attribution of observed differences to the conditioning method rather than to confounding factors such as data variability or training configuration.

\subsection{Dataset}
We use a curated subset of the Semantic Timbre Dataset \cite{cameron_vaetraindataset_2024,Cameron_SemanticTimbreDataset_2025}, comprising 1,771 monophonic electric guitar spectrograms. Each sample is labeled with:

\begin{itemize}
    \item One of the Semantic Timbre Dataset's 19 semantic timbre descriptors \cite{Cameron_SemanticTimbreDataset_2025}.
    \item One of four timbre descriptor magnitudes (25\%, 50\%, 75\%, 100\%), where timbre magnitude represents the magnitude of the Guitar Rig Pro 7 effect that corresponds to each relevant Semantic Timbre Dataset timbre descriptor \cite{Cameron_MPhilThesis_2024} applied to monophonic electric guitar notes recorded by Pedroza et al. \cite{egfxset_pedroza_2022}.
    \item One of 23 musical pitches ranging from E4 to D6.
\end{itemize}

All of this audio was preprocessed into STFTs computed with a 1024-point FFT and 512 hop size at 22.05kHz. These STFTs were then used to train the VAEs to obtain different latent spaces for comparison.

\subsection{The VAE Models}

\begin{figure}[h]
    \centering
    \includegraphics[width=\linewidth]{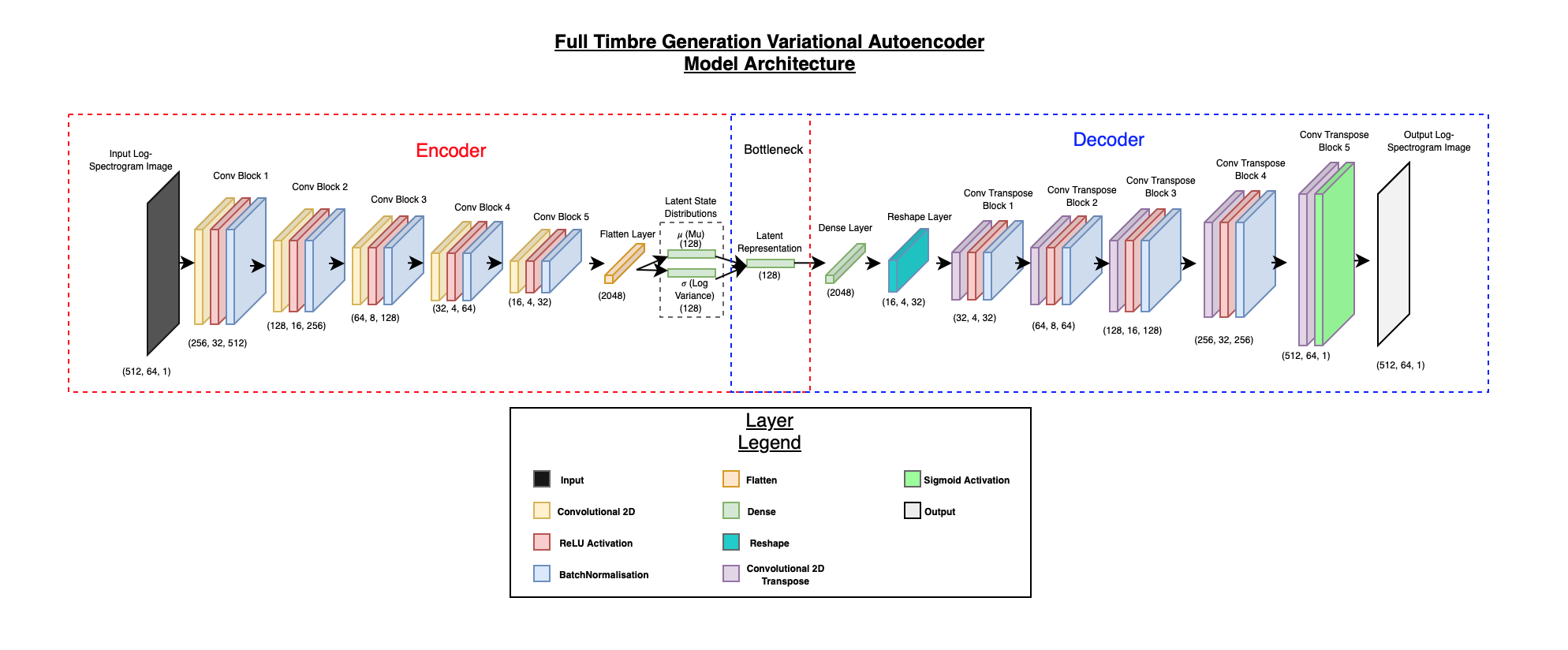}
    \caption{Semantic Timbre Generation VAE model architecture. The encoder maps input spectrograms to a 128D latent space. The decoder reconstructs log-magnitude spectrograms, which are converted back to audio.}
    \label{fig:FullVAEModelArchitecture}
\end{figure}

Three VAE models' latent spaces were compared. Each model uses the same convolutional encoder-decoder VAE architecture with a 128-dimensional latent space, which can be seen in Figure \ref{fig:FullVAEModelArchitecture}, and identical training hyperparameters. The models differ only in the conditioning strategy:

\begin{itemize}
    \item \textbf{Unsupervised VAE}: Learns representations solely from spectrograms with no auxiliary input.
    \item \textbf{Descriptor-Conditioned VAE (CVAE)}: Conditions on a 42-dimensional one-hot vector comprising: 19 binary descriptor indicators (multiplied by descriptor magnitude), and 23 one-hot pitch labels. The condition vector is concatenated to both encoder and decoder inputs.
    \item \textbf{Perceptual Feature-Conditioned VAE (AudioCommonsCVAE)}: Conditions on 7 continuous perceptual timbre features (brightness, depth, hardness, roughness, warmth, sharpness, boominess) from the AudioCommons models \cite{Pearce_AudioCommonsTimbreDescriptors_2017,Pearce_AudioCommonsTimbralModels_2019}, along with 23 one-hot pitch labels, for a 30-dimensional condition vector. Again, the condition vector is concatenated to both encoder and decoder inputs.
\end{itemize}

All models were trained for 300 epochs using the Adam optimizer \cite{kingma_adam:_2014} with a learning rate of 0.0005, a batch size of 64, and following the approach by Cameron \cite{Cameron_MPhilThesis_2024}. Training metrics and spectrogram normalization details are provided in our code repository \cite{Cameron_SemanticTimbreDataset_Code_2025}.

\subsection{Evaluation Metrics}
To evaluate the interpretability and structure of each latent space, we define a set of both standard and domain-specific metrics.

\subsubsection{Clustering Metrics}
\label{subsubsec:ClusteringMetrics}
Silhouette scores \cite{Rousseeuw_SilhouetteScore_1987} and cluster purity scores \cite{Manning_PurityScore_2008} were computed across all three VAE models for latent embeddings grouped by timbre descriptor. These metrics assess how well the latent space separates sounds produced via effects described by different timbre descriptor classes.

\subsubsection{Timbre Descriptor Compactness}
\label{subsubsec:TimbreDescriptorCompactness}
A `compactness' measure was developed that quantified how tightly each descriptor’s samples cluster across the entire latent space. Compactness was calculated using Euclidean distance as $Compactness = 1/(1 + \bar d)$ where $\bar d$ is the mean pairwise Euclidean distance between all samples of a given descriptor. Higher values indicate tighter clustering, with the transformation ensuring scores remain in the $[0,1]$ range.

\subsubsection{Individual Timbre Descriptor Magnitude Organization}
\label{subsubsec:TimbreDescriptorMagOrg}
Silhouette scores \cite{Rousseeuw_SilhouetteScore_1987} were computed for magnitude clusters occurring within each set of samples per timbre descriptor to analyze how well magnitude levels (25\%, 50\%, 75\%, 100\%) organize within each timbre descriptor across all pitches. This indicates whether increasing timbre magnitude corresponds to a smooth trajectory in the latent space.

\subsubsection{Pitch-Conditioned Analysis}
\label{subsubsec:PitchConditionedAnalysis}
Initial observations showed the latent spaces were organized primarily by pitch, with timbre as a secondary factor. This reflects the known dependence of timbre perception on pitch \cite{Grey1977TimbrePitch,Luo2019DisentangleTimbrePitch}. To avoid analyses being dominated by this pitch-first structure, specialised evaluations were introduced:
\begin{itemize}
    \item \textit{Within-Pitch Timbre Descriptor Separation}: Silhouette score \cite{Rousseeuw_SilhouetteScore_1987} analysis of timbre labels within each pitch cluster, measuring how distinctly timbre variations are encoded once pitch is held constant.
    \item \textit{Cross-Pitch Consistency}: A new `consistency' measure was developed that evaluates how consistently each descriptor-magnitude combination is positioned in the latent space across different pitches. $Consistency = 1/(1 + \sigma_{distances})$ where $\sigma_{distances}$ is the standard deviation of the descriptor-magnitude pairwise Euclidean distances. A high consistency value means the model encodes a timbral quality similarly regardless of pitch, which is a desirable property for a timbre-controlled generator.
\end{itemize}

\subsubsection{Timbre Magnitude Trajectory Metrics per Descriptor}
\label{subsubsec:TimbreMagTrajPerDescriptor}
We introduce two novel interpretability measures, designed to assess whether increasing timbre descriptor magnitudes correspond to predictable trajectories in a model's latent space:
\begin{enumerate}
    \item \textit{Linearity}: Linearity measures how straight the path is from minimum to maximum timbre magnitude in the latent space. It's calculated as the ratio of straight-line distance to actual path distance through intermediate magnitudes. Perfect linearity yields 1, while highly curved progressions approach 0.
    \item \textit{Step Consistency}: Step consistency measures the uniformity of distances between successive timbre magnitude levels. This indicates whether the latent space changes in a predictable, smooth manner as an effect described by a timbre descriptor is turned up. It's calculated using the coefficient of variation of consecutive distances ($CV = \sigma_{distances} / \mu_{distances}$, where $\sigma_{distances}$ = standard deviation of the consecutive distances and $\mu_{distances}$ = mean of the consecutive distances), transformed to a consistency score ($Consistency = 1 / (1 + CV)$) where higher values indicate more uniform progression steps.
\end{enumerate}


\section{Results}
\label{sec:Results}

\begin{table*}[h]
\caption{Evaluation metrics for each model. Bold values indicate the best score in each column.}
\label{tab:metrics}
\centering
\setlength{\tabcolsep}{3pt} 
\renewcommand{\arraystretch}{1.1} 
\begin{tabular}{lcccccccc}
\hline
\textbf{Model} & \textbf{Global Sil.} & \textbf{Purity} & \textbf{Compact.} & \textbf{Magn. Sil.} & \textbf{Within-Pitch Sil.} & \textbf{Cross-Pitch Cons.} & \textbf{Linearity} & \textbf{Step Cons.} \\
\hline
VAE & $-0.0302$ & $0.1016$ & $0.0525$ & $-0.0135$ & $0.0531$ & $0.3354$ & $\mathbf{0.5908}$ & $0.8671$ \\
CVAE & $-0.0384$ & $0.1090$ & $0.0530$ & $-0.0152$ & $0.0256$ & $0.2056$ & $0.5346$ & $0.8726$ \\
ACCVAE & $\mathbf{-0.0134}$ & $\mathbf{0.1383}$ & $\mathbf{0.0630}$ & $\mathbf{-0.0100}$ & $\mathbf{0.0777}$ & $\mathbf{0.3782}$ & $0.5703$ & $\mathbf{0.8754}$ \\
\hline
\end{tabular}
\end{table*}

Table \ref{tab:metrics} summarises the evaluation metrics for all three models. Across most measures, the \textbf{AudioCommonsCVAE} achieves the highest scores, indicating that perceptual conditioning leads to more structured and usable latent spaces.

\subsection{Global Clustering Performance}
As shown in Table \ref{tab:metrics}, all models exhibit negative global silhouette scores, indicating overlapping timbre clusters due to the strong influence of pitch on latent organization. However, the AudioCommonsCVAE attains the best Purity (0.1383) and least negative silhouette score (-0.0134), suggesting that continuous perceptual features improve class separability compared to both the VAE and CVAE.

\subsection{Descriptor Compactness \& Magnitude Organization}
The AudioCommonsCVAE yields the highest compactness (0.0630), representing a $\sim$20\% improvement over the other models. This indicates that timbre descriptors form tighter clusters in its latent space. Similarly, its magnitude silhouette score (-0.0100) is the least negative, implying slightly better organization of intensity levels within descriptors.

\subsection{Pitch-Conditioned Analyses}
When controlling for pitch, the AudioCommonsCVAE shows the strongest within-pitch separation (0.0777), outperforming the unsupervised VAE (0.0531) and the CVAE (0.0256). This reinforces the benefit of perceptually grounded features for encoding timbral variation independent of pitch. Cross-pitch consistency results reveal a substantial advantage for the AudioCommonsCVAE (0.3782) over the VAE (0.3354) and CVAE (0.2056), demonstrating better pitch-invariance, a key property for real-time control.

\subsection{Magnitude Trajectories}
All models achieve relatively high step consistency ($\approx$0.87), indicating predictable and uniform changes in timbre magnitude. Linearity scores are highest for the unsupervised VAE (0.5908) but remain similar across models, suggesting that explicit conditioning does not strongly influence trajectory straightness.



\section{Discussion}
\label{sec:Discussion}

Our results indicate that conditioning VAEs on continuous perceptual features yields clear advantages for structuring timbre latent spaces. While all three models exhibit a strong pitch-driven organization, a well-documented phenomenon in timbre perception \cite{Grey1977TimbrePitch}, the AudioCommonsCVAE achieves tighter descriptor clustering, greater pitch invariance, and more consistent magnitude organization than both the unsupervised VAE and descriptor-conditioned CVAE.

One-hot semantic labels, while aligning with common human terms and descriptors, are coarse and subjective. Two samples both labeled as `bright' may differ substantially in underlying acoustic content, introducing noise into the conditioning signal. Moreover, the discrete nature of one-hot conditioning offers no natural representation of intermediate timbral states, potentially fragmenting the latent space.

In contrast, the AudioCommons timbral perceptual features are continuous, physically grounded, and psychoacoustically validated \cite{Pearce_AudioCommonsTimbreDescriptors_2017,Pearce_AudioCommonsTimbralModels_2019}. This allows the model to exploit finer-grained correlations between conditioning input and spectral patterns, promoting smoother and more coherent latent trajectories.

\subsection{Implications for Interactive Timbre Control}
The high cross-pitch consistency of the AudioCommonsCVAE is particularly relevant for real-time control applications. In performance contexts, users expect a descriptor to have a consistent perceptual effect regardless of the note played. Without pitch invariance, identical control gestures could yield unpredictable results, reducing usability.

Additionally, all models exhibit high magnitude step consistency, meaning that gradual adjustments in descriptor intensity lead to predictable latent space movements. This is encouraging for continuous control interfaces such as sliders, gesture-based devices, or haptic controllers.

\subsection{Limitations}
The dominance of pitch in all latent spaces suggests that extended disentanglement strategies could further improve timbre separability. Moreover, the evaluation was limited to electric guitar timbres; generalising these findings to broader instrument families and more complex datasets remains an open question.

\subsection{Methodological Contribution}
Beyond model comparison, this work introduces a domain-specific evaluation framework for generative audio latent spaces. By combining standard clustering measures (see section \ref{subsubsec:ClusteringMetrics}) with our timbre-specific metrics (see sections \ref{subsubsec:TimbreDescriptorCompactness}, \ref{subsubsec:TimbreDescriptorMagOrg}, \ref{subsubsec:PitchConditionedAnalysis}, \ref{subsubsec:TimbreMagTrajPerDescriptor}), we provide tools for assessing not just separability, but interpretability of latent space structures, a key requirement for deployment in musical interfaces.


\section{Conclusion}
\label{sec:Conclusion}

We presented a comparative study of three VAE architectures for timbre modeling: an unsupervised VAE, a descriptor-conditioned VAE, and a perceptual feature-conditioned VAE. Using a curated electric guitar timbre dataset and a suite of both standard clustering and domain-specific interpretability metrics, we evaluated how each conditioning strategy shapes latent space organization.

Our results show that conditioning on continuous perceptual features produces a more compact, pitch-invariant, and descriptor-consistent latent space than either unsupervised learning or one-hot semantic conditioning. These properties are particularly valuable for interactive timbre control, where consistency and interpretability are essential. Beyond the model comparison, we contributed an evaluation framework for generative audio latent spaces that addresses controllability as well as separability.



\clearpage
\vfill\pagebreak

\bibliographystyle{IEEEbib}
\bibliography{refs}

@article{kingma_adam:_2014,
	title = {Adam: {A} {Method} for {Stochastic} {Optimization}},
	shorttitle = {Adam},
	url = {http://arxiv.org/abs/1412.6980},
	urldate = {2017-11-21},
	journal = {arXiv:1412.6980 [cs]},
	author = {Kingma, Diederik P. and Ba, Jimmy},
	month = dec,
	year = {2014},
	note = {arXiv: 1412.6980},
}

@misc{Cameron_SemanticTimbreDataset_2025,
	author       = {Joseph M. Cameron and Alan F. Blackwell},
	title        = {{The Semantic Timbre Dataset}},
	year         = {2025},
	howpublished = {\url{https://huggingface.co/datasets/JoeCameron1/SemanticTimbreDataset}},
	doi          = {10.57967/hf/5480},
	publisher    = {Hugging Face}
}

@inproceedings{Kingma_VAE_2014,
  author = {Kingma, Diederik P. and Welling, Max},
  booktitle = {{ICLR} 2014, Banff, AB, Canada, April 14-16, 2014, Conference Track Proceedings},
  eprint = {http://arxiv.org/abs/1312.6114v10},
  eprintclass = {stat.ML},
  eprinttype = {arXiv},
  title = {{Auto-Encoding Variational Bayes}},
  year = 2014
}

@article{peeters_descriptors_2011,
    author = {Peeters, Geoffroy and Giordano, Bruno L. and Susini, Patrick and Misdariis, Nicolas and McAdams, Stephen},
    title = "{The Timbre Toolbox: Extracting audio descriptors from musical signals}",
    journal = {The Journal of the Acoustical Society of America},
    volume = {130},
    number = {5},
    pages = {2902-2916},
    year = {2011},
    month = {11},
    issn = {0001-4966},
    doi = {10.1121/1.3642604},
    url = {https://doi.org/10.1121/1.3642604},
    eprint = {https://pubs.aip.org/asa/jasa/article-pdf/130/5/2902/15297399/2902\_1\_online.pdf},
}

@misc{egfxset_pedroza_2022,
  title={{EGFxSet: Electric guitar tones processed through real effects of distortion, modulation, delay and reverb}},
  author={Pedroza, Hegel and Meza, Gerardo and Roman, Iran R.},
  month={September},
  year={2022},
  publisher={Zenodo},
  version={Version 1.0},
  doi={10.5281/zenodo.7044411},
  howpublished={\url{https://doi.org/10.5281/zenodo.7044411}}
}

@misc{cameron_vaetraindataset_2024,
  author       = {Cameron, Joseph M.},
  title        = {{Monophonic Electric Guitar Notes to Train a Timbre Generation VAE}},
  month        = may,
  year         = 2024,
  publisher    = {Zenodo},
  version      = {1.0},
  doi          = {10.5281/zenodo.11398253},
  howpublished          = {\url{https://doi.org/10.5281/zenodo.11398253}}
}

@misc{Cameron_SemanticTimbreDataset_Code_2025,
  author       = {Cameron, Joseph M.},
  title        = {{SemanticTimbreDatasetCode GitHub Code Repository}},
  month        = may,
  year         = 2025,
  publisher    = {GitHub},
  version      = {1.0},
  howpublished = {\url{https://github.com/JoeCameron1/SemanticTimbreDatasetCode}}
}

@mastersthesis{Cameron_MPhilThesis_2024,
    author = {Joseph M. Cameron},
    title = {Recognising, generating, and
interpolating timbre in electric
guitars with semantic descriptors},
    school = {University of Cambridge},
    year = {2024},
    note = {https://github.com/JoeCameron1/MPhil-Dissertation-Project/}
}

@inproceedings{Esling2018GenerativeTimbreSpaces,
  author    = {Philippe Esling and Axel Chemla-Romeu-Santos and Adrien Bitton},
  title     = {Generative timbre spaces: Regularizing variational autoencoders with perceptual metrics},
  booktitle = {Proceedings of the 21st International Conference on Digital Audio Effects (DAFx)},
  year      = {2018},
  address   = {Aveiro, Portugal},
  url       = {http://dafx2018.web.ua.pt/papers/DAFx2018_paper_12.pdf}
}

@article{Caillon2020TimbreLatentSpace,
  author    = {Axel Caillon and Adrien Bitton and Philippe Esling},
  title     = {Timbre latent space: exploration and creative aspects},
  journal   = {arXiv preprint arXiv:2008.01370},
  year      = {2020},
  url       = {https://arxiv.org/abs/2008.01370}
}

@article{Yutani2024CVAEWavetable,
  author    = {Tsugumasa Yutani and Yuya Yamamoto and Shuyo Nakatani and Hiroko Terasawa},
  title     = {Wavetable Synthesis Using CVAE for Timbre Control Based on Semantic Label},
  journal   = {arXiv preprint arXiv:2410.18628},
  year      = {2024},
  url       = {https://arxiv.org/abs/2410.18628}
}

@article{Roche2021PerceptualTimbreVAE,
  author    = {Fanny Roche and Thomas Hueber and Ma\"{e}va Garnier and Samuel Limier and Laurent Girin},
  title     = {Make That Sound More Metallic: Towards a Perceptually Relevant Control of the Timbre of Synthesizer Sounds Using a Variational Autoencoder},
  journal   = {Transactions of the International Society for Music Information Retrieval (TISMIR)},
  volume    = {4},
  number    = {1},
  pages     = {52--66},
  year      = {2021},
  doi       = {10.5334/tismir.76}
}

@article{Natsiou2023InterpretableTimbreVAE,
  author    = {Anastasia Natsiou and Luca Longo and Sean O'Leary},
  title     = {Interpretable Timbre Synthesis using Variational Autoencoders Regularized on Timbre Descriptors},
  journal   = {arXiv preprint arXiv:2307.10283},
  year      = {2023},
  url       = {https://arxiv.org/abs/2307.10283}
}

@article{Pearce_AudioCommonsTimbreDescriptors_2017,
title = {Timbral attributes for sound effect library searching},
author = {Pearce, Andy and Brookes, Timothy and Mason, Russell},
journal = {AES E-Library},
pages = {2 - 2},
publisher = {Audio Engineering Society},
booktitle = {2017 AES International Conference on Semantic Audio},
address = {Erlangen, Germany},
year = {2017},
url = {https://openresearch.surrey.ac.uk/esploro/outputs/conferencePresentation/Timbral-attributes-for-sound-effect-library/99516677602346},
urldate = {2025-05-05}
}

@techreport{Pearce_AudioCommonsTimbralModels_2019,
       author = {Andy Pearce and Saeid Safavi and Tim Brookes and Russell Mason and Wenwu Wang and Mark Plumbley},
       title = {Deliverable D5.8: Release of timbral characterisation tools for semantically annotating non-musical content},
       institution = {University of Surrey},
       year = {2019},
       number = {D5.8}
}

@inproceedings{Bitton2019ModulatedVAE,
  author    = {Adrien Bitton and Philippe Esling and Tatsuya Harada},
  title     = {Modulated Variational Auto-Encoders for Many-to-Many Musical Timbre Transfer},
  booktitle = {ICLR 2019 Workshop on Deep Generative Models for Highly Structured Data},
  year      = {2019},
  url       = {https://openreview.net/forum?id=HJgOl3AqY7}
}

@inproceedings{Luo2019DisentangleTimbrePitch,
  author    = {Yin-Jyun Luo and Kat Agres and Dorien Herremans},
  title     = {Learning Disentangled Representations of Timbre and Pitch for Musical Instrument Sounds Using Gaussian Mixture Variational Autoencoders},
  booktitle = {Proceedings of the 20th International Society for Music Information Retrieval Conference (ISMIR)},
  year      = {2019},
  address   = {Delft, The Netherlands},
  pages     = {487--494},
  url       = {https://archives.ismir.net/ismir2019/paper/000091.pdf}
}

@article{Eerola2013MusicalCuesEmotion,
  author    = {Tuomas Eerola and Anders Friberg and Roberto Bresin},
  title     = {Emotional expression in music: Contribution, linearity, and additivity of primary musical cues},
  journal   = {Frontiers in Psychology},
  volume    = {4},
  pages     = {487},
  year      = {2013},
  doi       = {10.3389/fpsyg.2013.00487}
}

@article{Rousseeuw_SilhouetteScore_1987,
title = {Silhouettes: A graphical aid to the interpretation and validation of cluster analysis},
journal = {Journal of Computational and Applied Mathematics},
volume = {20},
pages = {53-65},
year = {1987},
issn = {0377-0427},
doi = {https://doi.org/10.1016/0377-0427(87)90125-7},
url = {https://www.sciencedirect.com/science/article/pii/0377042787901257},
author = {Peter J. Rousseeuw}
}

@incollection{Manning_PurityScore_2008,
place={Cambridge},
title={Flat clustering},
booktitle={Introduction to Information Retrieval},
publisher={Cambridge University Press},
author={Manning, Christopher D. and Raghavan, Prabhakar and Sch\"utze, Hinrich},
year={2008},
pages={321–345}
}

@article{Grey1977TimbrePitch,
  author    = {John M. Grey},
  title     = {Multidimensional perceptual scaling of musical timbres},
  journal   = {Journal of the Acoustical Society of America},
  volume    = {61},
  number    = {5},
  pages     = {1270--1277},
  year      = {1977},
  doi       = {10.1121/1.381428}
}

\end{document}